\documentclass[10pt]{article}
\usepackage{graphicx} 
 \usepackage{xcolor}
\usepackage{ulem}
\usepackage{epsfig}
\usepackage{latexsym}
\usepackage{amsmath}
\usepackage{hyperref}
\usepackage{bigints}
\usepackage {eucal}

\begin{document}

\title{\Large \bf 
A  theory of best choice selection \\ through objective arguments \\ grounded in Linear Response Theory concepts.
 }

\author{ \large \bf  Marcel Ausloos \textsuperscript{1,2,3,4}\\  \bf Giulia Rotundo \textsuperscript{5} \\ \bf Roy Cerqueti \textsuperscript{6,7} \\
 \\\\ \\
$^1$ School of Business,
University of Leicester,    Brookfield, \\ Leicester, LE2 1RQ, UK    \\$e$-$mail$ $address$:
ma683@le.ac.uk \\
$^2$ GRAPES,   rue de la Belle Jardini\`ere 483, \\B-4031, Angleur, Li\`ege, Belgium  
 \\$e$-$mail$ $address$:
marcel.ausloos@uliege.be\\
$^3$ Department of Statistics and Econometrics, \\The Bucharest University of Economic Studies,  \\ Caeia Dorobantilor 15-17,  010552 Bucharest, Romania \\$e$-$mail$ $address$:marcel.ausloos@csie.ase.ro \\
 $^4$    Universitatea Babeș-Blolyai, \\ Str. Mihail Kogălniceanu nr. 1, 400084, Cluj-Napoca,  Romania
 \\ 
$^5$ Sapienza University of Rome, Department of Statistical Sciences, \\  Piazzale Aldo Moro 5,  I-00185, Rome, Italy
 \\$e$-$mail$ $address$: giulia.rotundo@uniroma1.it
\\
$^6$ Sapienza University of Rome, Department of Social and Economic Sciences, \\  Piazzale Aldo Moro 5,  I-00185, Rome, Italy
  \\$e$-$mail$ $address$: roy.cerqueti@uniroma1.it
\\
$^7$ University of Angers, GRANEM -- SFR Confluences, \\F-49000 Angers, France
\\
}
\maketitle
\newpage
  \begin{abstract}
In this paper, we propose   
how to use objective arguments grounded in statistical mechanics concepts in order to obtain a single number, obtained after aggregation,  which would allow to rank "agents", "opinions", ..., all defined in a very broad sense.  We aim toward any process which should  a priori demand or lead to some consensus in order to attain the presumably best choice among  many possibilities. In order to precise the framework, we discuss previous attempts, recalling trivial "means of scores",  - weighted or not, Condorcet paradox, TOPSIS, etc. We demonstrate through geometrical arguments on a toy example, with 4 criteria,   that the pre-selected order of criteria  in previous attempts makes a difference on the final result. However, it might be unjustified. Thus, we base our "best choice theory"  on  the linear response theory in statistical mechanics:   we indicate that one should be calculating correlations functions between all possible choice evaluations,  thereby avoiding an arbitrarily ordered set of criteria.  We justify the point through an example with 6 possible criteria. Applications in many fields are suggested.   Beside,  two  toy models serving as practical examples and illustrative arguments are given in an Appendix.
 
 \end{abstract}

\newpage

\section{Introduction}
 
Statistical physics has found widely opened research topics outside its classical aims  in economics and sociology nowadays \cite{stauffer2000grand,stauffer2004introduction,Savoiu12,kutner2019econophysics,grech2020simplicity}. Thus, consider the interplay between sociology and physics: sociophysics.

Forget Hobbes, Quetelet, Comte, Verhulst, ....,  there should be no need to point out, as an introduction or justification of this paper, that Galam is  our contemporary pioneer of modern sociophysics \cite{galam1982sociophysics,galam2008sociophysics}. Due to space limits and the content of  papers in this special  issue, let us only refer to the relevant specific papers of interest for our consideration here below \cite{galam97Ising,galam2004contrarian,galam2013modeling,galam2014isitnecessary,galam2016invisible,galam2020tipping,galam2022opinion,biondi2012formation}.

 One of Galam's goals is to set a framework, provide techniques, and search for conclusions on the dynamics of opinion formation in various  societies. His work leads to finding conditions for consensus (some sort of "equilibrium"), chaotic states, and intermediary complex phases in multi dimension diagrams. However, it is  not obvious that the findings pertain to what common people would refer to or call "the best choice".  But what is "the best choice"? It should be admitted that "the  best choice" is a highly relative state or concept.  The answer contains both ingredients from both personal ("selfish") and global ("self-effacing")  points of view. Moreover, the dependence on exogenous and endogenous conditions is huge, but this is left for political discussions elsewhere. 

We consider that the final states in Galam's studies, or models, and subsequent studies by many, result from a too heavily weighted stochastic set of constraints, or hypotheses. Surely, opinion dynamics  result from individual "votes" (that means, choices) due to herding or because of contrarians \cite{galam2004contrarian,galam2016invisible,dhesiausloos2016modellingsoliton}.  Nevertheless,  does the dynamics implies a good choice, or worse, is the  choice (that means, vote) the best choice? That seems to be a crucial point, not only  at elections times in  democracies, but also in choices like (voluntarily limiting references to a few papers) 
 in media \cite{Lehmann71}, 
including music genre-fication  networks \cite{LambiotteAusloosEPJB},
 in economy \cite{Camagnietal15}, 
  including regional studies \cite{novac2020dynamic} and drawdown market prices sizes at  
 speculative  times \cite{rotundo2007maximumdrawdown},
 in academia,  including  scholarly journals ranking 
 \cite{journalclassranking},     research networks clustering \cite{cerquetimattera2023clustering}, world universities ranking  \cite{LiuCheng05,Florian07} or samba schools ranking \cite{sambaschoolsstatista,sambaschoolstaylor}, and
 in sport \cite{JACC94,ausloos2014primacyUEFA,Malcataetal14,AORFiccadenti2022}, 
  etc.

More explicitly in the academic domain,  what is the best choice when hiring or promoting a colleague, when appointing a vice-chancellor,   when selecting teams for research grants?  In the sport domain, the ranking of football  teams  or of cyclist racing teams are obtained through apparently objective numbers, but the rules can often (or even always) be debated and challenged \cite{AusloosEJOR,AusloosentropyShannon}. The same remark holds for the Nobel prize, the Pulitzer prize, the Goncourt prize, Oscars or Cesars awards which are given through highly subjective, not objective, criteria; not discounting facing a choice
 between roads going from X to Y, for going on holidays or to a restaurant,  .... ? What is the best equipment or car  or cell phone to buy?  What is the best food, from a health point of view? All questions mixing subjective and objective criteria.

This boils into the fundamental and practical question : What are the criteria needed for reaching the best choice?   In other words, how should one  conclusively rank a set of "things",  "people" or "teams" in a constrained set of criteria?

This leads to remember that somewhat the "final choice" leads to a paradoxical situation, the Condorcet paradox \cite{Condorcet1785,YoungLevenglick1978,Young1988}    for example\footnote{The  Condorcet method   \cite{Condorcet1785,YoungLevenglick1978,Young1988}  is a voting system that will always select the candidate whom voters prefer to each other candidate, when compared between  them one at a time.    }.
 Further considerations on comparing "preferred choices" lead to  Arrow's  incompatibility theorem  \cite{arrow1950difficulty}. Moreover, the order of criteria might lead to ambiguities \cite{HeDeng2023SoftComp}.
 
The discussion, and the subsequent answers,   should pertain to a comparison of the evaluation methods, according to criteria  \cite{krawczyk2019heider,krawczyk2021structural}.  
Most of them are based on previous achievements, even though their forecasting value, not mentioning consistency  for future achievements or impact are far from certain.

The drastically annoying deduction seems to stem from the plethora of "parameters", i.e., possible criteria. Practically, one turns toward aggregation processes \cite{[13],Munda2012,[60]}, going  from multi dimensions toward a single number. That makes life pretty difficult when one turns 
 toward modelling.  Thus, one wishes to have some indubitable argument; often that  means to have rigorous mathematical theorems. However, this is often hard to implement  in particular for laymen (or lay women, or lay others). Therefore,  mathematical arguments might be by-passed through physics concepts which allow metaphors and analogies, like in modern statistical mechanics, ...  like in Galam's view of social thought dynamics, say on networks.

  Therefore, after outlining  elements for discussion from information theory, arguments of geometry, and surely arguments on complex systems and sociophysics, we propose a powerful  argument grounded in linear response theory (LRT): in LRT   the coefficients (magnetic susceptibility, transport coefficients, etc.),  measured in laboratory and theoretically discussed, are defined through the correlations between the (fluctuations of the) dependent variables. Thus, we suggest to calculate all correlation functions implying the relevant variables in the sociophysics research topics of interest, in particular when searching for opinion formation, dear to Prof. Galam. That idea seems to be  missing in previous work. We claim that it is leading to a more objective hierarchy of values in  opinion formation choice, etc., topics.

Within this set of considerations, a study framework can be conceived  together with applications,  thereby leading to the following structural content of this paper.

Sect. \ref{IandM} contains a brief review of "old" and "modern" techniques for selecting, and ranking, agents or events, like the rank-size "laws". We discuss such attempts, recalling trivial "means of scores",  be they weighted or not, Condorcet paradox, etc. We recall  preference aggregation techniques, like the "Maximum Likelihood Rule" (Sect. \ref{MLR}),  TOPSIS (Sect. \ref{TOPSIS}), and  a few other preference ordering approaches (Sect. \ref{otherpref}).

However, the aggregation problems have two different aspects:  rank aggregation or score aggregation. They are briefly distinguished in Sect. \ref{scoreagg}.

In particular, in Sect. \ref{SoC},  we  start from  
 geometrical aspects for ranking scores, or  measures, and for later aggregating multi-valued measures . 
Furthermore, we demonstrate through such geometrical arguments  that the order of criteria might make a difference on the final result, but might be unjustified. It is easily observed that if only 3 criteria are used, the procedure leads to an undebatable result.  We show on a set of 4  criteria for a "toy model" that the order is  drastically relevant and influences the outcome  after aggregation.

We base our "best choice theory" in Sect. \ref{ProbandSol}, upon the linear response theory.  We indicate that one should be calculating correlations functions between all possible choice evaluations. 
 Appendix  A recalls the fundamentals of "linear response theory' in statistical physics.

We conclude  \textcolor{black}{ with some emphasis,  if it is needed, that the best ranking is obtained from a method based on rigorous arguments in obtaining a final score and ranking.  We are aware that subjective arguments often influence the final decision.}
We suggest applications  in a few domains in Sect. \ref{conclusions}. Two examples are found in Appendix B.

\section{Modern Ideas  and  Methods }\label{IandM}

In this Section, we consider ideas and methods  about ranking and selection of the best outcome due to inequalities between events, agents, also called "variables", thereby assuming that a hierarchy takes places and leads to the best choice \cite{Pumain}. 

Before recalling modern ideas, let us remind the reader that the selection process leads to some  so called rank-size  (RS) display or tables. Indeed, the  rank-size analysis is the basic way of measuring disorder in a population: the largest "size" gets the first rank, and a hierarchy, whence an ordering through inequality,  is deduced \cite{Pumain}.  It is a hierarchy description. {\it In fine}, this leads to considering whether empirical laws follow patterns, thereby suggesting models. 

The RS law derives from the  analytical form presented by the variables, usually  ranked in descending order as a function of the (discrete) index $i$ giving a "rank" to each of the  \textcolor{black}{ $g_i$, $i=1, \dots, N$ variables. }
 The cumulative \textcolor{black}{  law $\sum_i\;g_i$ leads to } the  "cumulative concentration distribution curve"(CCDC).    \textcolor{black}{ The  sum $\sum^N_i\;g_i$ over all the $i$ elements can serve as a normalisation value. One easily obtains the "normalised CCDC"}. When the  \textcolor{black}{  normalised} CCDC  goes over   \textcolor{black}{ a}  80\%  threshold, it defines the Pareto rank $r_P$. The "Pareto   \textcolor{black}{principle}" expects this rank $r_P$ to be  equal to $N/5$. Thus, the RS  method is convenient, and "more meaningful",  for large populations, i.e., when the size or/and rank  ranges can be large.  This is rarely the case when only a few selection criteria are implied.

   For completeness, nevertheless, let us mention that  Marfels  \cite{Marfels71} distinguishes several types of concentration ratios, according to weighting schemes and  their structure which can be discrete or cumulative \cite{BikkerHaff2002}. Beside such ratios, inequality aspects are often discussed in order to touch upon  socio-economics concerns \cite{Cowell11}.

        Furthermore, we can recall that when the goal is to find a compromise between the various rankings, the statistical median,   is thought to be the most appropriate solution \cite{Young1995,Garciaetal2014}.   However,  in situations  in which  decision making should be a way of compromising between conflicting decisions, the  "Maximum Likelihood Rule", discussed in the following section, Sect.\ref{MLR},  makes sense.

        \subsection{Maximum Likelihood Rule} \label{MLR}
 Recall that a choice demands some ranking. This is usually done by combining ordered preference lists into a single consensus   value, as in a reviewing procedure \cite{Garciaetal2014,Torres2005}. We outline the  "Maximum Likelihood Rule" (MLR), that is based directly on rankings, not on the scores.

  In brief, methods of preference aggregation, as the MLR,  are based on the concept of   pairwise preference notions \cite{LeCamMLR90}, much used in economics and in  opinion formation along the simple majority voting rule, - without discussion on the scores. 
 It is also known as the Kemeny  \cite{Kemeny1995} rule    (or the Kemeny-Young method \cite{Young1995}).  In order to go beyond the limit of the method, one has  introduced a variant  taking into account the behavioural argument of "Blindness to Small Changes" (BSC) \cite{VarelaRotundo}.

 Mathematically, a MLR ordering is defined as one that minimises the total number of discrepancies among all the reviewers in their \underline{pairwise preferences} between all options.  It can be viewed as a voting scheme that determines not just a single chosen winner, but an entire ordered list. Therefore, the MLR ordering satisfies, generally, the most possible reviewers as regards their stated rankings of options. It does not use any information about how much higher a candidate is ranked over another, but only  a relative ordering\footnote{   It can be worth to  point out that  MLR is a Condorcet method. }.
     
     Practically, one can say that the method counts the   \underline{pairwise preferences}  and applies over each of them the majority rule. 
     
    Examples abound on  applying this rule and finding "solutions" \cite{Lambiotteetal2007}. 
   One short illustration is given in Appendix   \textcolor{black}{ B.} 
                         
    The second step of the method is counting the number of times in which one candidate is ranked over another. In so doing,  the MLR  emphasises that the  first-ranked choice   wins against all other options in individual pairwise comparison.
    Similarly, the MLR second ranked  choice would win against all other options (except the first ranked), and so on.

             In order to complete this brief subsection, we may consider that  the MLR implies that candidates having the same score are considered to belong to the same "indifference set". 
              If this occurs, one decides to  list the "candidates" (specifying the "agents" or "events") one after the other, e.g., by "alphabetical order", in such a set. One may think about other ways to manage equal scoring, but there is no need to foresee any special treatment here to take care of  {\it ex aequo} positions in the ranking.   Nevertheless, in most evaluation procedures, a linear order is requested for the final ranking; the most annoying point is when two "agents" are {\it ex aequo}   on the first rank. Usually, in order to achieve such a goal, or resolve such a dilemma, an extra information is {\it a posteriori}  added for discriminating equally scored candidates.  That might be unfair, but this "solution" is left for more legal considerations, outside the present paper.
             
\subsection{Technique for Order Preference by Similarity to Ideal Solution } \label{TOPSIS}

A "Technique for Order Preference by Similarity to Ideal Solution" (TOPSIS)  
was originally developed by  Hwang and Yoon \cite{[1]}, with further \textcolor{black}{pertinent}  developments  \cite{[2],[3],Lai1994Topsis},  
 \textcolor{black}{e.g., see Parida et al. \cite{Paridaetal2017} and Chakraborty et al. \cite{TopsisChakraborty2024}}
 is a method of compensatory aggregation that compares a set of alternatives by identifying weights for each criterion, normalising scores for each criterion and calculating the geometric distance between each alternative and the ideal alternative, which is the best score in each criterion.

TOPSIS continues on working attractively  over various application territories; see \url{https://en.wikipedia.org/wiki/TOPSIS}.

\subsection{Score Rather than Rank Aggregation}\label{scoreagg}

However, some fundamental emphasis between processes in the preferential ordering  problems must be made through  distinguishing rank aggregation  from score aggregation.  Often, the ranks are not known,  but are deduced from scores. However, as pointed out, one does not always know how the  scores are obtained (recall the number of stars in the Michelin guide or the Gault  \&  Millau scores, of restaurants, \textcolor{black}{ or }  
to the ranking order in Trip Advisor!). In such cases, scores are incompatible  and sometimes  incomprehensible;  ranking only  makes some sense, allowing for {\it ex aequo}s. 
 
 However if scores are known, from a reliable set of measures  and for a given set of criteria, some objective construction can follow, leading to some meaningful score aggregation. A scoring function can be defined as $f^{(X_j)}(s_i)$, with $s_1, s_2,s_3, ..., s_m$, being the scores  on each $i$-th criterion, for  a given  "agent" $X_j$, with $j=1, ..., n$.
 
 The aggregation results, from  the set of $n$ "events, agents" and the $m$ scoring criteria, sorted out for example  in decreasing order, lead to  finding the $top-k$ "candidates",  according to the scoring function  imposing $ f([s_i])<f([s_i'])$.
 The objective is clear: to compute the "$top-k$" "agents" with the minimum cost.
While the target is clear, the methods can be quite different, depending on the choice of $f$.
The very first problem stems in  the sorting out algorithm.  Let us point out to the "Fagin algorithm" \cite{Fagin1974}, to the "medrank algorithm" \cite{medrank},
and  to the "threshold algorithm" \cite{threshold}.
For some completeness, another method  
is the "Borda count" \cite{Borda}: for each ranking, one assigns  a score $S$  equal to the number of objects it defeats. The total weight of $S$ is the number of points it accumulates from all rankings.

However, these rank or score ranking methods can be challenged because they are missing a key ingredient,   i.e., the selection order of the criteria.

\subsection{On the Sequence of Criteria. A Geometrical Perspective}\label{SoC}

 In brief, many (all?) previous works ignore fairness through a crucial factor: the sequence of criteria. It can be pointed out and easily illustrated that an ascertainment of criteria has an effect on ranking values; this is of common knowledge to anyone having participated in  surveys \cite{AlwinKrosnick,Sanchez92} and/or selection processes.
 
 Suppose that there are 3 criteria giving a numerical value ($a$, $b$, and $c$, respectively) for the agent or event.  One can define a coordinate system with three axes stemming from some origin $O$, in equivalent directions (such that the angle between each axis is therefore $2\pi/3$) and  plot the values on each "criterion axis". Next, one way to aggregate the  3 values is to consider them as 3 sides of a triangle, and calculate the triangle area  which becomes the "score".
 
 In case, it would be necessary, one may recall that 
in order to find the area of a triangle with 3 sides, one uses the Heron's formula: indeed, the area of a triangle ($S$) with 3  known sides $a$, $b$, and $c$ is calculated from
 
\begin{equation} \label{EqS}
  S =  \sqrt{[s(s-a)(s-b)(s-c)]}\;,
 \end{equation}
   where $s$ is the semi-perimeter of the triangle,  i.e.,  $s = (a + b + c)/2$.
     
On the other hand, knowing two sides $b$ and $c$ around  an angle $A$, 
the formula to calculate the area of a triangle is given by $(1/2) \times b\times c \times sin(A)$. Obviously, 
\begin{equation}  \label{Eq2_4axes}
S=b \times c/2
\;\;\;\;\;\; if  \;\;\;\;\; A=\pi/2, 
 \end{equation}
 while 
\begin{equation}  \label{Eq3_6axes}
S= (\sqrt 3/4)\times b\times c
\;\;\;\;\;\; if  \;\;\;\;\;  A=2\pi/3  \;\;\;\;\;(or \;\; \pi/3).
 \end{equation}
 
In the latter case, that means that whatever the order of criteria, the  total area, i.e., the sum of the three triangles areas,  is invariant and equal to $S= (\sqrt 3/4)\times (b\times c + a \times b + a \times c)$.

\begin{figure}
\centering 
\includegraphics[height=8cm]{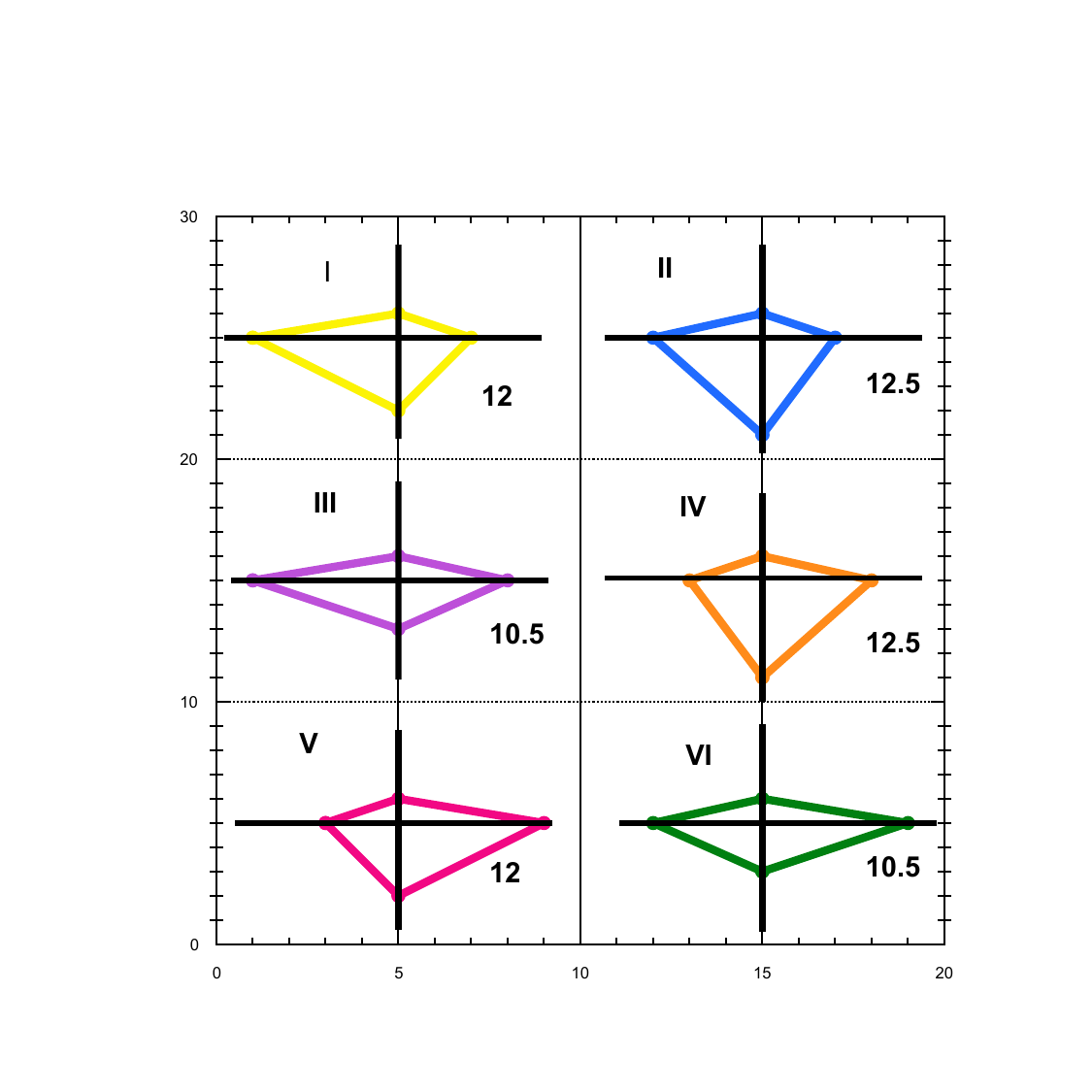}
\caption{Demonstration that starting from 4 criteria the shape of the polygon based on  the variable values leads to different areas; the toy values are $a=1,b=2,c=3,d=4$, leading to 6 possible polygons, but 3 different sizes; the areas values are given in arbitrary units.
}\label{Plot_1_6squares}
\end{figure}

However the matter is different when there are more than 3 criteria, or sustaining axes. Consider the case of 4 criteria. Let the axes be forming a coordinate system with 4 axes in symmetric directions. The permutation of axes leads to 6 different polygons. A toy example, say with  $a=1$, $b=2$,  $c=3$, $d=4$,  can be seen on Fig. \ref{Plot_1_6squares}.
 The 4 inner  triangles in each polygon are rectangular triangles for which each area is easily obtained (the total area is given in the Figure,   in arbitrary units); it can be easily noticed that by symmetry only 3 different areas are relevant.   Therefore, this leads to 3 different sizes, if the area is considered to be the aggregated number for ranking the  agents or events.

  When the number of criteria increases, whence many more polygons can be drawn, the area of such polygons can always be decomposed into a number of triangles, for which each area can be calculated from Eq. (\ref{EqS}), 
remembering that if  the length of two sides ($a$ and $b$) of the  known angle ($C$) between them, in any triangle, one can easily calculate the third side length of the triangle through Al-Kashi formula,  i.e.,
  \begin{equation} \label{3trdside}
  c^2\; =\; a^2 \;+ \;b^2 -2\times a \times b \times cos(A)
  \end{equation}

\begin{figure} 
    \includegraphics[width=0.5\textwidth] {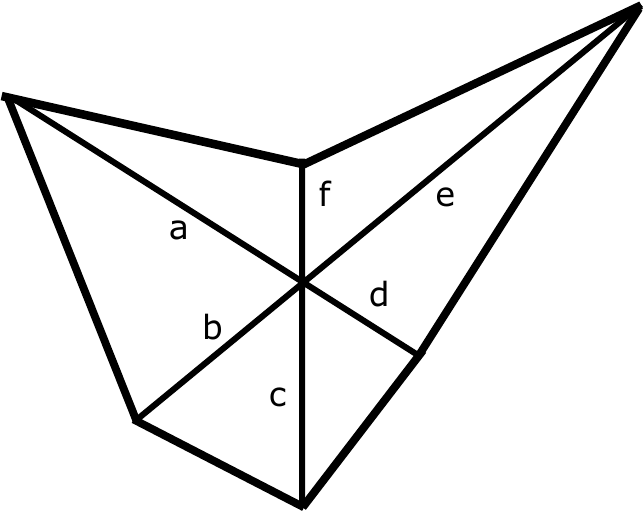}
    \includegraphics[width=0.5\textwidth] {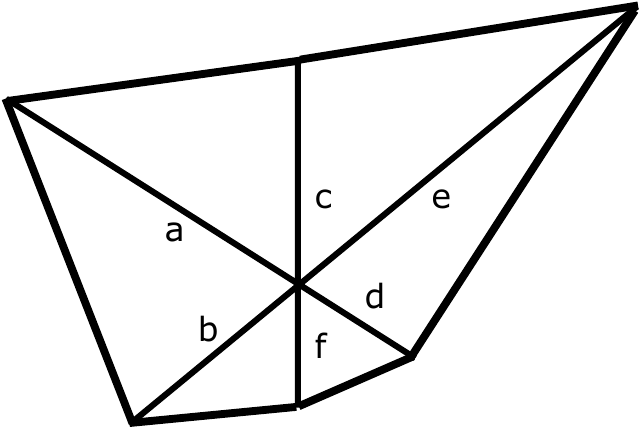}
\caption{A toy case of six criteria is presented in two sub-figures. The variables take values $a,b,c,d,e,f$ 
 in both displays.  Notice the disposition of the axes, whose arbitrariness concerns the discussion ground of the present study. A different disposition in the relative order of the axes leads to a different area for the hexagon.} \label{fig1}
\end{figure}

    \section{The problem and its solution} \label{ProbandSol}
Thus, the way to proceed    goes as follows.  First, take a survey with as many possible criteria which are needed for a population whatever its size.  One may suppose, without loss of generality, that the choice is based on Likert scales, whatever the   useful range $[0,5]$, $[0,7]$, $\dots$, $[0,20]$.  

 Notice that it could be of interest to consider the mean and standard deviation of the data distribution on each axis, such that one can obtain a possible "universal comparison" of  final rankings and hierarchies.
If the statistical characteristics make sense, the measured values could be normalised or, at least, constrained to vary in an identical range.  One could assume that all the values involved in the decision process belong to $[0,1]$.
The value 0 is achieved in the worst situation for the considered parameter (bad realisation) and, accordingly, 1 is taken in the "best choice" case.   Nevertheless, such a scaling is not a fundamental request. Although of interest, the normalisation  aspect is not further discussed and is left for further work.

Then,   one reports the values on equidistant axes. These are disposed in a regular star-shaped network with a central node, the common starting point and identical angles between consecutive axes; the angle value obviously depends on the number of  axes (criteria). The non-common nodes of the consecutive segments are connected. The resulting polygon  has a number of sides identical to the number of  evaluation parameters; the polygon is not necessarily regular; It can be considered made of triangles. If $N$ is the number of axes, the largest possible number of different polygons is $[(N-1)!/2]$. The number of different triangles is  $N$!. 

In this paper, let us focus on hexagons, since evaluation processes are very often associated to six different parameters.  

In Fig. \ref{fig1}, we present an example for two hexagons  out of 60 possible cases; let  $a,b,c,d,e,f$ (e.g., $ \in [0,1]$), the values of the six parameters. The surface of the resulting hexagon $\mathcal{S}$ can be easily computed.  
The evaluation score coincides with the value of $\mathcal{S}$, so that high $\mathcal{S}$ means high score. However, notice that $\mathcal{S}$ belongs to the range   associated to the limiting cases $a=b=c=d=e=f=0$ (worst
result of the evaluation process, with $\mathcal{S}=0$) and $a=b=c=d=e=f=1$ (best result of the evaluation exercise, with
$\mathcal{S}=\frac{3\sqrt{3}}{2}$).

Clearly, the surface of the hexagons shown in Fig. \ref{fig1} depends on how the segments are disposed: 
 the values of the six measures are always $a,b,c,d,e,f $  
 but the placement of $c$ and $f$ have been reverted  in  both displays.  

Let $\mathcal{S}_1$ and $\mathcal{S}_2$ be  the surface of the hexagons in Fig. \ref{fig1}, (left) and (right),
 respectively.  We have
$$
\mathcal{S}_1=\frac{\sin{(\pi/3)}}{2}\left[ab+bc+cd+de+ef+fa\right]=\frac{\sqrt{3}}{4}\left[ab+bc+cd+de+ef+fa\right],
$$
$$
\mathcal{S}_2=\frac{\sin{(\pi/3)}}{2}\left[ab+bf+fd+de+ec+ca\right]=\frac{\sqrt{3}}{4}\left[ab+bf+fd+de+ec+ca\right];
$$
 in the  case  $b+d-e-a \neq 0$, one has that $
\mathcal{S}_1 =  \mathcal{S}_2$, if and only if $c = f$.

Such a remark leads to the questionability of the evaluation exercise through polygons. Indeed, 
the resulting polygons,  differently shaped according to the disposition of the axes,   have different surfaces. This means that the evaluation is strongly dependent on how   selection parameters are included in the graphical representation of the problem, as we do stress again.

To overcome such an inconsistency,   whence in order to obtain a more indubitable scoring,
we propose to take the average $\bar{\mathcal{S}}(a,b,c,d,e,f)$ of all the  possible polygon surfaces, thus over all the possible dispositions of the axes,   - here corresponding to 6 evaluation criteria:
\begin{equation}\label{averageS_Eq5}
\bar{\mathcal{S}}(a,b,c,d,e,f)=\frac{\sin{(\pi/3)}}{
5!}\sum_{\mathbf{H}(a,b,c,d,e,f)}\left[x_1 x_2+x_2 x_3+x_3 x_4+x_4
x_5+x_5 x_6+x_6 x_1\right],
\end{equation}
where
$$
\mathbf{H}(a,b,c,d,e,f)=(x_1, \dots, x_6) \in
\{a,b,c,d,e,f\}^{(6)} : x_i \neq x_j, \text{ for each } i \neq j.
$$

By construction, $0 \leq \bar{\mathcal{S}}(a,b,c,d,e,f) \leq \frac{3\sqrt{3}}{2}$.   Such a mean surface  measure effectively does the intended job, i.e.,  providing a consistent score of the evaluated object whose six parameters (criteria) take values $a,b,c,d,e,f$.

This is consistent with the measure concept resulting into calculating the correlation between fluctuations in    statistical physics, as  found in
 linear response theory: 
 see Appendix    A. Examples of Applications are found in Appendix B.

\bigskip
 
 \section{Conclusions} \label{conclusions}
    
    In brief, the present paper aims at resolving a paradoxical situation when searching for an objective ranking procedure and subsequently observing some  hierarchy through a set of criteria for some population.
    This   sort of  scientometrics practice is tied  to considerations found in geometry and statistical physics, complementing aspects of research on opinion dynamics, in Galam's numerous studies.

Observe the technical and somewhat philosophical  frames, in the present ranking  study.  Decisions often concern rather small sub-systems,  like, the funding of a project, the evaluation of a research group or of that an individual,  at his or her hiring or promotion. Statistical laws cannot be immediately applied for such cases in which,  due to the (very finite) size of the system,  fluctuations which are not  necessarily  due to stochastic causes play a crucial role. Indeed, recall that  scientific work  is part of a social system; its actors are human beings. Moreover, statistics are 
like snapshots and do not  often allow to  predict the future very well! In practice, indeed, any newly introduced criterion or  "measure"  will be met by  the capacity of humans to  interact and making decisions.
 
Without going back to the Bible and raising the question who is the just and who will be saved, - according to a single judge, it can be considered that the modern time of scientific reasoning on choice, - when several partners and judges are implied, goes back to     Arrow \cite{arrow1950difficulty}. He considered the preference aggregation problem, that is the problem of passing from a set of known individual preferences to a pattern of social decision making. His now  often quoted  theorem (with various wordings)  has  nevertheless shown that the  difficulties  met in the building  process of preference aggregation are  very general. The theorem implies that         no rank order voting system can convert the ranked preferences of individuals into a community-wide (complete and transitive) ranking,   beside also meeting a specific set of natural criteria.

 In fact,  
  why is it  impossible to reach a choice?  
  Usually, one constructs several filters  and {\it a priori} decides on the order of their applications,
  liked in the decision tree (DT). That leads to a discussion on ranking the filters rather than the candidates.
  
   \textcolor{black}{ Indeed,  in the "decision tree" (DT) scheme,  due to the order of filtering criteria,  the  final choice is much biased. The order of filters is often adapted {\it a priori\;}  in order to select the final choice, e.g.,  a candidate, or maintain candidates in competition with others during the selection process, for hypocritical, political, or other reasons. 
   }
   
   In opinionology,  \textcolor{black}{ the selection }   process consists in  projecting from a multidimensional space onto various planes, and finally finding the intersection of the distribution \footnote{ In physics, it is like projecting on various "external field axes" or making a scalar product.}. There might not be any "solution".  That  further may mean that the set (of "candidates", "events", etc.), to be ordered, is either not to be ranked or that the filters are not appropriate. 
   
Thus, we  propose, starting from Galam and other socio-physicists considerations on the dynamics of choices, a procedure for obtaining the "best choice"  through an objective    statistical physics method\footnote{  \textcolor{black}{It seems fair to argue with a comment of a reviewer through this footnote. This allows to emphasise peer reviewer contributions. The reviewer claims (not an exact quotation) that our {\it method is not better or worse than another; it is merely another type of classification.Moreover it will not be applied by the Hollywood Academy to award Oscar prizes.} First of all, we are not using the word "better". We emphasise that our method is more rigorous, contains less arbitrariness, and is based (through an analogy) on major statistical physics concepts.We might regret that our method might not be applied In Hollywood. Tongue in cheek, we admit that we are not aware of all criteria used by the Academy in order to award Oscar's. The same holds true in other branches of opinion formation. In a decision process the final ranking might be due to  hidden or specifically weighted criteria. Our method does not apply  outside the objective world. We do not consider subjective criteria. }}. 

Of course, the situation might not be closed, since the choice of criteria is left for many discussions between agents, be they surveyees or surveyors,  or other stakeholders.

 Nevertheless, our arguing, after outlining elements of reflexion from information theory, geometry, surely   complex systems, as found in sociophysics considerations, proposes a strong argument for calculating choice values. It is grounded in linear response theory : in the latter, as recalled in Appendix A, coefficients to be measured in laboratories and theoretically discussed, are defined through the correlations between the dependent variables; thus, we suggest to do the same in  sociophysics, - to calculate all correlation functions implying the relevant variables, thereby avoiding an arbitrarily ordered set of criteria. In particular, this should be useful and meaningful in the modelling of opinion formation, processes dear to Prof. Galam.  That methodology is missing in sociophysics previous work. Thus, such a LRT basis, rather easily implemented, should  be leading to a more objective hierarchy of values ahead of selection processes.

\bigskip

    \bigskip

\bigskip \newpage
\bigskip

{\bf    Appendix A. Linear Response Theory}\label{LRT}

\bigskip 
\bigskip 
The linear response theory  (LRT) was independently  invented by Green \cite{Green52,Green54} and Kubo \cite{Kubo66}: it describes the coefficients relating the effect of a perturbation on a thermodynamic system in equilibrium.
The LRT has given a general proof of the fluctuation-dissipation theorem which states that the linear response of a given system to an external perturbation is expressed in terms of  the\underline{ correlations between fluctuations} properties of the system in thermal equilibrium \cite{AndrieuxGaspard04}. 

   Kubo  \cite{Kubo66}  considered the application of a magnetic field to an equilibrium system, and demonstrated that the magnetic susceptibility can be defined through the average of the correlations between the magnetic moment density fluctuations. The LRT    also well applies to the description of the electrical conductivity or the thermal conductivity \cite{Mori1965,Ausloos78}, even if the system is characterised as being in  a non-equilibrium state.

Consider a perturbation   $B(t)$, e.g. a magnetic field,  and  search for  the response of the system  $M(t)$, i.e., the magnetisation. Usually one demands to obtain  $M(t)=\chi B(t)$.

By analogy,     \textcolor{black}{ in opinion formation and related rankings, one may consider that the "response" is going to be the final score. It is extracted through the correlations between fluctuations of the values attributed to "candidates" (in a wide sense, usually called  "agents", - which might not be humans)  through the various  criteria. The mere fact that criteria are introduced for later agent ranking. is considered to induce "perturbations" in the original system.  The external field is the exogenous decision of introducing criteria.
 }

 \textcolor{black}{ Subsequently, } one may consider that the external  (field) perturbation  is  applied at  some time $t_0$ to some (say) agents in some population;  the "population system" is in so doing moved away from its equilibrium, and is characterised through a non-equilibrium ensemble average. Thus, this   leads to the measured  score of each agent as a result of interactions  in response to the opinion field perturbation,   - as in most of Galam's models.

In practice, considering a weak interaction with some external field, one can obtain the resulting score  by performing an expansion in powers of the perturbation \cite{Green52,Green54,Kubo66,AndrieuxGaspard04,Mori1965,Ausloos78}.
The leading term in this expansion is independent of the field, but the next term  describes the deviation from the equilibrium behaviour in terms of a linear dependence on the external perturbation through the correlations between the system fluctuations. The average score is the linear response function, i.e.,  the quantity that contains the (microscopic) information on the system and how it responds to the   perturbing field. 
 
    \newpage

    {\bf    Appendix B. Two examples}\label{2examples}
    
\bigskip 
\bigskip 
    Among the many possible examples, in. the various research fields outlined in the Introduction and Conclusion sections, we have selected two examples which are likely of interest to many readers and researchers in physics, particularly in sociophysics. The first example concerns the promotion of researchers. The second example pertains to the evaluation of football (soccer) players. The former stresses how to reach a final ranking from ranks obtained through various filters. Three methods, outlined in the main text, are so compared. The latter example stresses that a meaningful ranking can be obtained from a final aggregation number based on measured values, analysed through correlations as proposed in the main text..
    
    \begin{itemize}
    \item Example 1. Researcher promotion.
    
    Consider for some illustration that 5 researchers (A, B, C, D, E) are applying for promotion.  The relevant selection committee has decided to base its ranking of the value of the candidates on 4 criteria,   related to the  impact factor (IF) of the journal where a paper has been published\footnote{The 5 candidates were required to submit their best  10 papers from such an IF point of view.}. The committee calculates
      \begin{itemize}
     \item  the IF mean on the best 5 papers: $\overline{B5}$;
     \item  the iF mean on the worst 5 papers: $\overline{W5}$;
     \item  the IF mean on the oldest 5 papers: $\overline{F5}$;
     \item the IF mean on the most recent 5 papers: $\overline{L5}$.
     \end{itemize}  
  The timing of the publication is considered irrelevant at this stage.   The resulting mean values and the corresponding candidate rank is given in Table \ref{summarytable12x} for each criterion\footnote {The raw data is available on demand.}.

 Let the committee be  (i) admitting that the ranks are more relevant than the scores, and (ii) the ranking of candidates be  in descending order of  the criteria scores. 
 
  (i)  It is immediately seen that 
   candidate A has the highest $\overline{B5}$ (= 7.8) and $\overline{F5}$ (=7.8); author E has the maximum score on the last 5 ($\overline{L5}$) (= 7.60), but author B is not reaching any highest score, among these competitors, under any criterion.    We may also remark that in this toy example, due to the nature of the evaluation, the  candidate D lays at the bottom on each ranking, no matter   the specific criterion is used.
     A difficulty arises in the need of  ranking   the 4 others, since each of them is a winner for at least one committee member (or criterion):  A   and C  even winning twice,   but on different criteria of course; B and E are twice {\it ex aequo}, but not near the top places.
     Thus, we have shown that  different criteria lead to different rankings, but the more so the toy model implies that there is no obvious  final choice  as was indeed codified by Arrow's theorem. Moreover, there is no indubitable hierarchy, along this simple aggregation process.
  
  (ii) Next, consider the Maximum Likelihood Rule (Sect. \ref{MLR}) method based on rankings, not on the scores, as given in Table \ref{tablescores}).  Recall that  a MLR ordering is defined as one that minimises the total number of discrepancies among all the criteria in a pairwise preference scheme. The MLR ordering   does not use any information about how much higher a candidate is ranked over another, but only  a relative ordering. 
     Practically, the method counts the pairwise preferences. The second step of the method consists in counting the number of times in which one candidate is ranked over another (from Table  \ref{tablescores}).  In so doing,  the MLR    emphasises that the  first-ranked choice   wins against all other options in individual pairwise comparisons.
    Similarly, the MLR second ranked  choice would win against all other options (except the first ranked), and so on.    The present case outcome is reported in Table \ref{tablePreference}.
 
  It  can  be  observed  from  Table \ref{tablePreference}  that D has the lowest  number of potential preferences for promotion, whence  is properly ranked as last. It can be remarked that the coalition of criteria formed only by  
   $\overline{B5}$ and $\overline{L5}$ can be a decisive one on the ranking of E before B. However the rank relations between A and C are not so neat.
    Therefore,  the selection committee would be faced with its self-application of Arrow's theorem consequences, again.
    
  (iii)   Finally, let us consider the newly proposed method.  It boils down to calculating the surfaces of rectangular triangles and its subsequent averaging, - here for 3 types of polygons  with 4 sides. Using Eq.(\ref{Eq2_4axes}), one obtains the results displayed in Table \ref{T3}.  
  It seems obvious that this ranking makes sense. Thereafter, one may conclude that the proposed method is more justified and advantageous than classical ones.

              \begin{table}\label{summarytable12x}
       \begin{center} 
       \begin{tabular}{|c||c|c||c|c||c|c||c|c|c|c|c|c|c|c|c|c|c|  }
     \hline &\multicolumn{2}{|c||}{$\overline{B5}$ }&\multicolumn{2}{|c||}{$\overline{W5}$ }&\multicolumn{2}{|c||}{$\overline{F5}$ }&\multicolumn{2}{|c|}{$\overline{L5}$}  \\ \hline \hline
      A	& \underline{7.80}	&1&3.20&2&\underline{7.80}&1&3.20&4	  \\	
      B	& 7.00&4&2.20&3&2.20&4&	7.00&2  \\	
      C& 7.60&2&\underline{3.60}&1&7.60&2&3.60&3	 \\	
      D&  6.40	&5&1.00&5&5.60&3& 1.80&5 \\	   
      E	& 7.60	&2&2.20&3&2.20&4&\underline{7.60}&1 \\ 	\hline 
 	   \end{tabular} 
       \end{center}
      \caption{  
      Profile of 5 candidates  (A, B, C, D, E) having published 10 papers,  listed in chronological order,   with the candidate evaluation through various criteria, proposed by  examiners;  the best score  is \underline{underlined} for each criterion;  the corresponding  rank is given; notations as  in the text.
       }\label{tablescores}
     \end{table}
         
       \begin{table}
                      \begin{center} 
                      \begin{tabular}{|c||c|c|c|c|c|c|c|c|c|c|c  }
                      \hline  
                        &    A& B& C & D& E  \\	\hline \hline 
                     A  &	$-$ &	3  & 2   &  4 & 3  \\	
                     B  &	1   & $-$  & 1   &  3 & ((0))	  \\	
                     C  &	2   &	3  &$-$  &  4 & (2)	 \\	
                     D  &	0   &	1  & 0   & $-$& 1 \\	   
                     E  &	1   &	((2))  & (1)   & 3  & $-$ \\ 	\hline 
          	   \end{tabular} 
                      \end{center}
                     \caption{  MLR application: Table reporting how many $a_{i,j}$ times, according to Table \ref {tablescores},
                    the candidate in row $i$ is ranked before the candidate in column  $j$; parentheses indicate {\it ex aequo}s. 
                      }\label{tablePreference}
                    \end{table}     
     
              \begin{table}\label{T3}
       \begin{center} 
       \begin{tabular}{|c||c|c|c||c||c||c|c||c|c|c|c| }
     \hline &\multicolumn{3}{|c||}{$area \times 2$ } &$total$& $average$& rank\\ \hline \hline 
      A	& 12&15&15&42&14	&1  \\	
      B	& 40&42&42&124&41.33&4  \\	
      C& 16&15&15&46&15.33&2 \\	
      D&  80&80&80&240&80&5 \\	   
      E	& 24&25&21&70&23.33&3 \\ 	\hline 
 	   \end{tabular} 
       \end{center}
      \caption{  Table reporting how  5 candidates  (A, B, C, D, E) are ranked  according to the proposed method; $area$ is the surface of each relevant 4 sided polygon; $total$ is the sum of the area of such polygons; $average$ is the average area.
       }\label{T3}
     \end{table}

  \newpage
     \item Example 2. Football (soccer) players.
  
     When this paper writing was being completed, we came across  some related application which we are pleased to relate.  It is about football players, ranking them on their "market value", "potential", salaries, age, and many statistical "criteria": see $https://sofifa.com/players$;  or for teams:  \url{https://sofifa.com/teams}. Another example of "player value" can be found on internet sites like  \url{https:https://www.laliga.com/en-GB/player/robert-lewandowski} 
     on which 8 criteria are displayed\footnote{ two axes, "yellow cards" and "red cards", might have 
 a value=0, reducing the display to a hexagonal pattern}. Radar picture comparisons can be created: 
  \url{https://www.laliga.com/en-GB/comparator/players?player1=witsel}.
 
      In particular, e-players can form their own team, based on real players,  considering six criteria, displayed on  
      a regular hexagon thereafter transformed into a colourful pattern; e.g., see 
     \url{https://sofifa.com/player/183277/eden-hazard/230036/}.
     
      The six skill measures, obtained through several sub-criteria, are given in Table \ref{Hazard}.
      In brief, SHO: SHOOTING: determines finishing skill and shot power, including penalty success;
      PAS: PASSING: denotes ability to successfully pass the ball with vision;
      DRI:  DRIBBLING: denotes ball control, agility and balance;   
      DEF: DEFENDING: notes tackling and interceptions;, 
      PHY: PHYSICALITY: notes strength and stamina,  and aggressiveness;
      PAC: PACE:  notes the speed and the acceleration of the player.  
      
       Notice that it is not obvious to us how the "overall note"    is obtained. Since only one hexagon is shown, it is clearly one for which the axes are {\it a priori} chosen, whence are not conforming to our recommendations about a tentative ranking (of  football players) upon some unique value which is resulting from some aggregation process.

             \begin{table}\label{Hazard}
       \begin{center} 
       \begin{tabular}{|c||c||c|c|c|c|c|c|c|c| }    \hline  
                        $"measure"$& skill:& SHO& PAS& DRI& DEF & PHY& PAC  \\	\hline \hline 												 77	&	SHO	&	-	&	6237	&	6545	&	2695	&	4774	&	6083		\\			
81	&	PAS	&	6237	&	-	&	6885	&	2835	&	5022	&	6399		\\			
85	&	DRI	&	6545	&	6885	&	-	&	2975	&	5270	&	6715		\\			
35	&	DEF	&	2695	&	2835	&	2975	&	-	&	2170	&	2765		\\			
62	&	PHY	&	4774	&	5022	&	5270	&	2170	&	-	&	4898		\\			
79	&	PAC	&	6083	&	6399	&	6715	&	2765	&	4898	&	-		\\					\hline 
 	   \end{tabular} 
       \end{center}
      \caption{  The 6 skill variables considered for measuring the "value" of E. Hazard, with their respective measure according to  
     \url{https://sofifa.com/player/183277/eden-hazard/230036/}; the data in the matrix corresponds to the area (divided by $ \frac{\sin{(\pi/3)}}{2}$, see Eq.(\ref{Eq3_6axes})) of the 15 possible triangles, - to be arranged in 60 different hexagons.
       }\label{T3Hazard}
     \end{table}

\begin{table}[!ht]
\centering 
  \begin{tabular}{|c||c|c|c|c|c|c|c|c|c|c|c| }    \hline    
   & $\nu$ & min. & Max.& Total area& mean &  Std Dev & skewness & kurtosis \\
\hline \hline
$sofifa$&	6&2170	&	6885&	29248 & 4874.7 &	1912.2 &	-0.4457	&	-1.4123	\\
LRT&	60&28298	&	29600&	1734432 & 28907 &	437.04 &	0.3076	&	-1.5345	\\
\hline 
\end{tabular}
\caption{ Main statistical characteristics of the  distribution of (i) top line:  the 6 triangle areas forming the single hexagon in  the $sofifa$ website,  (ii) bottom line: the 60 possible hexagons  areas characterising E. Hazard along the  6 skill axes, following the LRT method described in the text.
}
\label{Tablestats}
\end{table}

 Thereafter, in order to avoid useless decimals, we measure the area of triangles, whence of corresponding hexagons  in $\sqrt{3}/4$ units; see Eq.(\ref{Eq3_6axes}). 
      
      Consider one case in order to describe how the resulting aggregation player value is obtained, ahead of some selection process; consider E. Hazard.  According to the mentioned web site,  
     \url{https://sofifa.com/player/183277/eden-hazard/230036/}, his        
     "best overall  = 82".  However, the average of his skill values equals 69.83, as easily obtained from the values given in 
     Table \ref{T3Hazard} first column. Moreover, the latter Table reports the data corresponding to the area (divided by $ \frac{\sin{(\pi/3)}}{2}$, see Eq.(\ref{Eq3_6axes})) of the 15 possible triangles.   For space saving, not all 60 relevant hexagons can be displayed, nor the whole list of their areas.  
 For some completeness, the statistical characteristics of the polygon areas distribution is given in Table \ref{Tablestats}. The first line reports the characteristics for the distribution  of triangle areas on the diagram displayed on 
     \url{https://sofifa.com/player/183277/eden-hazard/230036/}; it corresponds to a mere $\nu=6$ axes diagram; the second line details the distribution of the $\nu=60$ hexagon areas considered along the present method. It is remarkable that  the LRT result shows a much narrower distribution, thus a more convincing statistics. The sign of the skewness is also in favour of the LRT method. 
     
     Not considering here a comparison with other players, we should point that such a comparison  demands a final aggregation score. In order to have a universal rule, we suggest that the best is to measure the "player hexagon area(s)" with respect to the perfect player, i.e., the largest regular hexagon.  For the $sofifa$  and the LRT case, one obtains 0.4875 and 0.4818, respectively.
     
    \end{itemize}
    
    \bigskip
    
    As a conclusion of this Appendix, notice that we have first debated on how to reach a hierarchical selection, through three methods, using numerical examples. The illustration is based on a  toy case made of a set of 5 candidates applying for promotion through a 4 criteria selection process. Next, we have presented the whole scheme resulting in the aggregation score for a given individual examined through 6 values or criteria, along two methods. 
    
    We have stressed the objective advantages of the LRT based method in both examples.
    
    Therefore, we can conclude from both toy cases here above, it seems to us, that the usability and the advantage of the LRT  methodology both bring much support to considerations in the sociophysics interdisciplinary field searching for objective ranking, whence  leading to a justified choice. We emphasise that the main  justification is  the use of correlation functions, as in LRT, i.e., a pertinent statistical physics theory basis.

          \end{document}